\documentclass[%
 aip,
 amsmath,amssymb,
 reprint,%
]{revtex4-2}

\usepackage{graphicx}% Include figure files
\usepackage{dcolumn}% Align table columns on decimal point
\usepackage{bm}% bold math
\usepackage[utf8]{inputenc}
\usepackage[T1]{fontenc}
\usepackage{mathptmx}
\usepackage{etoolbox}

\makeatletter
\def\@email#1#2{%
 \endgroup
 \patchcmd{\titleblock@produce}
  {\frontmatter@RRAPformat}
  {\frontmatter@RRAPformat{\produce@RRAP{*#1\href{mailto:#2}{#2}}}\frontmatter@RRAPformat}
  {}{}
}%
\makeatother
\begin{document}

\preprint{AIP/123-QED}

\title{Multiple feedback based wavefront shaping method\\ to retrieve hidden signal}

\author{Nazifa Rumman*}
\affiliation{Electrical, Computer, and Systems Engineering Department, Rensselaer Polytechnic Institute,Troy, NY, USA}
\author{Tianhong Wang}%
\author{Kaitlin Jennings}
\author{Pascal Bass\`{e}ne}
\author{Finn Buldt}
\affiliation{Department of Physics, Applied Physics and Astronomy, Rensselaer Polytechnic Institute, Troy, NY, USA}
\author{ Moussa N’Gom}
\affiliation{Department of Physics, Applied Physics and Astronomy, Rensselaer Polytechnic Institute, Troy, NY, USA}
\affiliation{Center for Ultrafast Optical Sciences,
University of Michigan, Ann Arbor, Michigan 48109}
\email[Corresponding author: ]{rumman@rpi.edu}
\date{\today}

\begin{abstract}
We present an optical wavefront shaping approach that allows tracking and localization of signal hidden inside or behind a scattering medium. The method combines traditional feedback based wavefront shaping together with a switch function, controlled by two different signals. A simple, in transmission imaging system is used with two detectors: one monitors the speckle signature and the other tracks the fully hidden signal (e.g., fluorescent beads). The algorithm initially finds the optimal incident wavefront to maximize light transmission to generate a focus in the scattering medium. This modulation process redirects the scattered input signal inducing instantaneous changes in both monitored signals, which in turn locates the hidden objects. Once the response from the hidden target becomes distinct, the algorithm switches to use this signal as the feedback. We provide experimental demonstrations as a proof of concept of our approach. Potential applications of our method include extracting information from biological samples and developing non-invasive diagnosis methods.
\end{abstract}

\maketitle

\section{Introduction}
Turbid materials including biological tissue or human skin scatter light and create complex interference patterns, known as speckle. This prevents light from being focused inside biological materials or retrieving information from objects behind scattering media, posing a fundamental limitation to many information processing and visualization technologies. Biomedical applications such as photodynamic therapy (PDT) \cite{PDT2011}  and optogenetics applications  \cite{opto2011} require directing light through skin or extracting information from target cells. These applications are either constraint to the outer layer of skin or become invasive through the insertion of an optical fiber \cite{optfiber2013}. \\
Wavefront shaping techniques using spatial light modulators (SLM) have emerged as an attractive proposition to address these issues \cite{park2018,YU2015}. The SLM modulates the incident wavefront in order to focus light through a scattering medium \cite{vel2007}, using methodologies based on phase conjugation \cite{yaqoob2008,Papadopoulos2012,Liu17}, feedback based optimization \cite{Stockbridge12,Conkey12}, the transmission matrix, \cite{TMgigan2010,NGom2018} or semi-definite programming \cite{Ngom2017} approach. The objective of most studies is to demonstrate focusing through scattering medium by utilizing speckle feedback e.g., speckle correlations feedback \cite{Stern2019}
rather than recovering signals/information hidden by a scattering medium. Nonlinear fluorescent signal has also been used as feedback to obtain diffraction-limited focusing through turbid samples \cite{Katz2014}. A shaped focus can be utilized to image masked objects with the help of memory effect, however, this does not  work for thick scattering samples \cite{Hsieh10,Daniel19,Yang14}. Approaches that only employ fluorescence feedback are often limited by the low intensity of the feedback signal. Many of these experiments must have the fluorescent feedback above a threshold signal to noise ratio (SNR) before the optimization process is initiated \cite{Boniface19}.
 
In this paper, we present a new approach that allows tracking, localization, and optimization of fluorescent signal emitted by objects hidden within and/or behind a scattering medium. We combine the traditional principle of feedback scheme based wavefront shaping together with a switch function that utilizes both the speckle intensity and the masked signal. The feedback signals are characterized by two different detectors: the first measures the speckle intensity from the highly scattering medium and the second detector monitors the signal generated by the hidden targets. The algorithm exploits the fact that modulating the incident wavefront to generate a focus in the speckle, redirects the scattered input fields. These redirected light fields will simultaneously induce a response from objects hidden within or behind the scattering medium, thereby locating targets. Once the target signal is located or reaches a threshold intensity, the feedback mechanism now switches to use the signal from the hidden object, then proceeds to enhance or optimize it. Our method is not dependent on a high SNR, or memory effect. This technique is a stride towards identifying and imaging hidden objects within highly scattering media. This approach will continue to improve biomedical imaging. More specifically, it can be an ideal biomedical tool for non–destructive, non–invasive diagnosis.

\section{principle}
\begin{figure}[b]
    \centering % <-- added
  \includegraphics[width=1\linewidth]{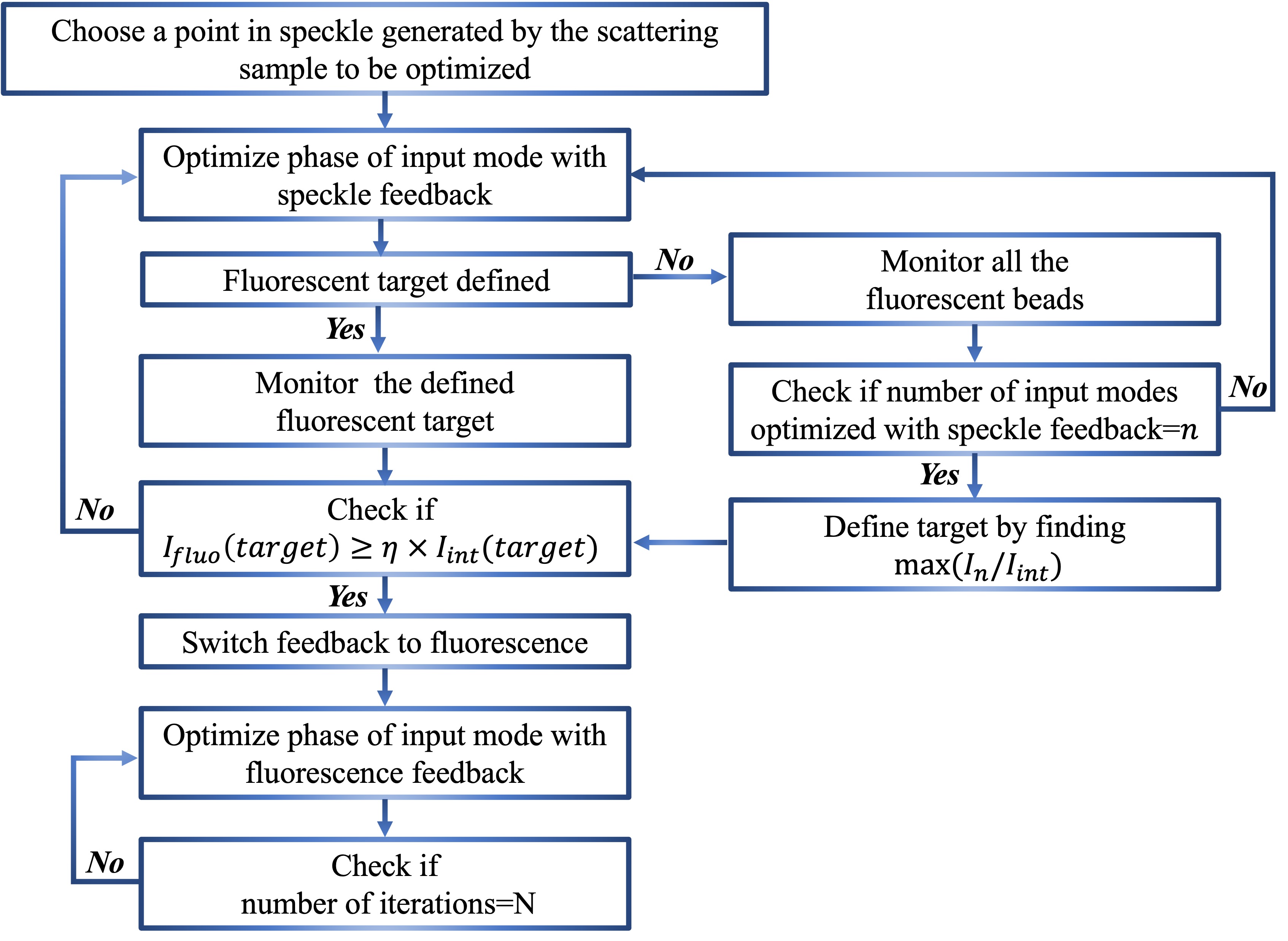}
\caption{A block diagram showing the steps of the switching feedback algorithm: The optimization process is initiated with speckle feedback. If the fluorescent target is predefined (case $i$), only that target is monitored, whereas in case $ii$, the algorithm finds the target after a preset number of iterations. In both cases, the switching condition has to be met to utilize fluorescence as the feedback. After switching feedback, the subsequent superpixels are modulated to optimize the fluorescent signal.}
\label{fig:flowchart}
\end{figure}

The algorithm is designed to find and probe a target deep inside a highly scattering medium. We start with the traditional iterative method utilizing continuous sequential algorithm \cite{vel2008} but a switch function is added. The algorithm starts with finding the optimal wavefront for creating a focus at the detector that monitors the speckle intensity. The pixels of the SLM are subdivided into N groups of pixels, known as superpixels. The phase of each superpixel is iterated from $0$ to $2\pi$ to find the phase that maximizes the feedback signal. These initial optimization processes are taken to generate a focus. The continuous variation due to the phase modulation can increase the signal from the hidden objects. After a predetermined number of iterations ($n$), e.g. at minimum the first row of superpixels  must be optimized; the algorithm starts to ‘look’ for this hidden signal and locates the buried target using the second detector. Once the signal from the target reaches a threshold intensity, the feedback signal is switched to come from the second detector. The algorithm now proceeds to enhance the signal from the target. The localization and enhancement of the target signal can be categorized into two cases:\\
$i$) We predetermine the target position where fluorescence could be expected. The iterative algorithm is then started to generate a focus at the chosen location using the speckle as feedback. The algorithm switches feedback when the fluorescence detected is above a defined threshold and then proceeds to enhance the signal.\\
$ii$) The algorithm finds the signal: the iterative optimization is started to generate a focus at a random speckle point. The algorithm independently monitors fluorescence signal from different targets. After modulating $n$ number of superpixels with speckle feedback the current fluorescence intensity($I_{n}$) is compared to the initial fluorescence intensity ($I_{int}$). The hidden fluorescent source that is emitting maximum intensity at this point of the optimization process is located and chosen as the target. The algorithm now switches to use the fluorescence as feedback to enhance its signal (see Figure \ref{fig:flowchart}).

\begin{figure}[b]
    \centering % <-- added
  \includegraphics[width=.75\linewidth]{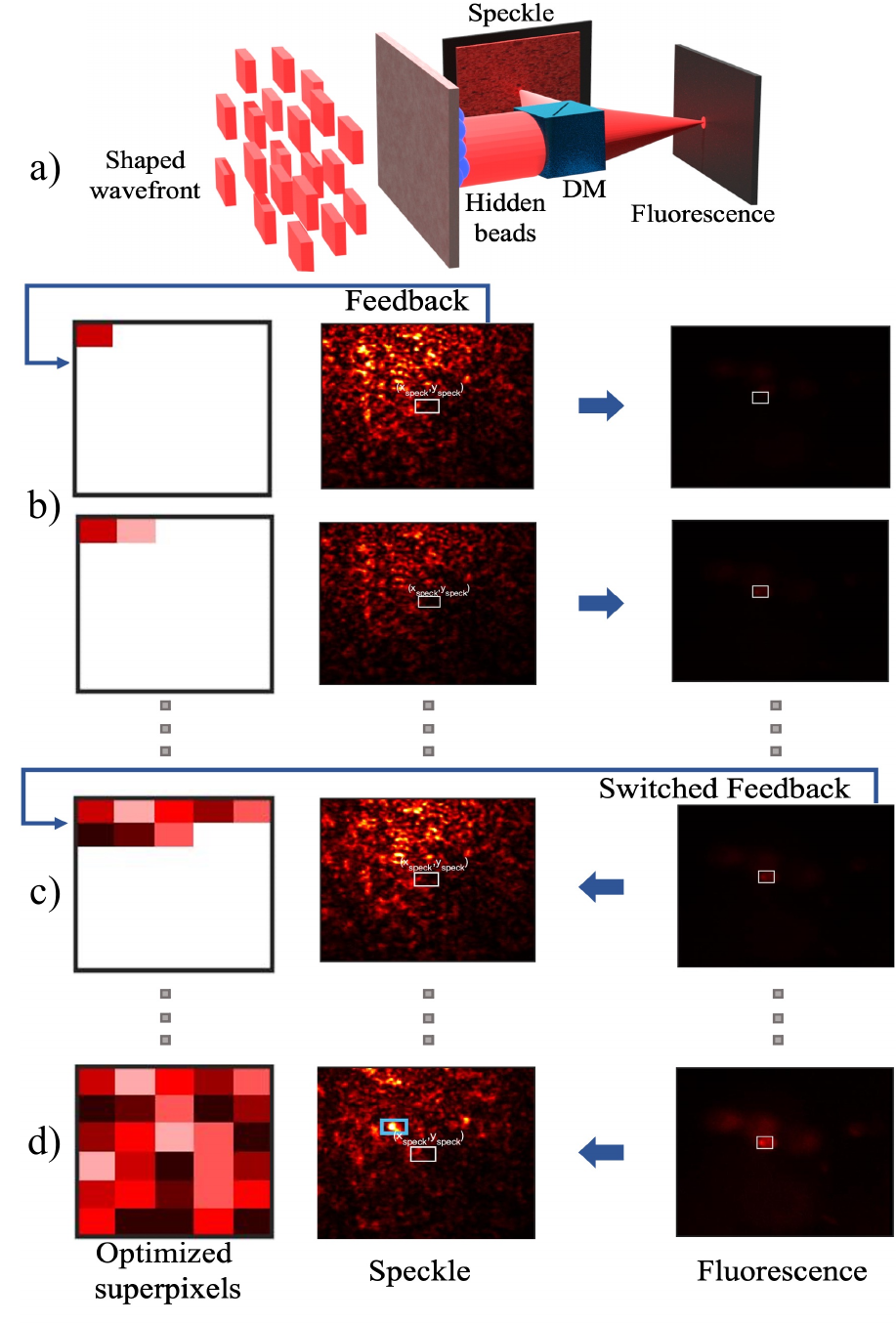}
\caption{a) Fluorescent beads hidden by the scattering medium are excited by the speckle and optimized with wavefront shaping. b) The fluorescence intensity detected initially is very low, so the phase of each SLM superpixel is modulated to maximize the speckle feedback, which induces variations in the monitored fluorescent signal at the same time. c) Once the fluorescent target has become distinct or gained enough intensity to perform the optimization process, the algorithm switches to update the remaining superpixels with fluorescence feedback d) Optimized fluorescence along with a focus in the speckle (marked by blue rectangle) can be obtained.} 
\label{fig:principle}
\end{figure}

For both of these cases, a seamless switch from the speckle to the fluorescence feedback is permitted, once the hidden signal’s intensity at the target position ($I_{fluo}(target)$) attains a value of at least $\eta$ times higher than the initial intensity ($I_{int}(target)$): $I_{fluo}(target) \geq \eta \times I_{int}(target)$, where $1.1 \leq \eta \leq 2$. The switching parameter, $\eta$, is both sample and case dependent. If the value of $\eta$ is too small the algorithm will switch feedback before the target intensity, $I_{fluo}(target)$ is sufficiently higher than the noise. In such a case the algorithm will not converge. If $\eta$ is too high the algorithm will not switch its feedback and the algorithm will proceed to generate a focus at the chosen speckle point\cite{Vel2015}. Once the switching conditions are fulfilled, the algorithm proceeds to adjust the phases of the remaining superpixels to enhance the masked signal. The speckle is not used as the feedback for the remainder of the process and is only monitored. A focus spot in the speckle will also be obtained as a result of the optimization process as shown in Figure \ref{fig:principle}.
\section{Experimental setup}
\begin{figure}[t]
    \centering
  \includegraphics[width=1\linewidth]{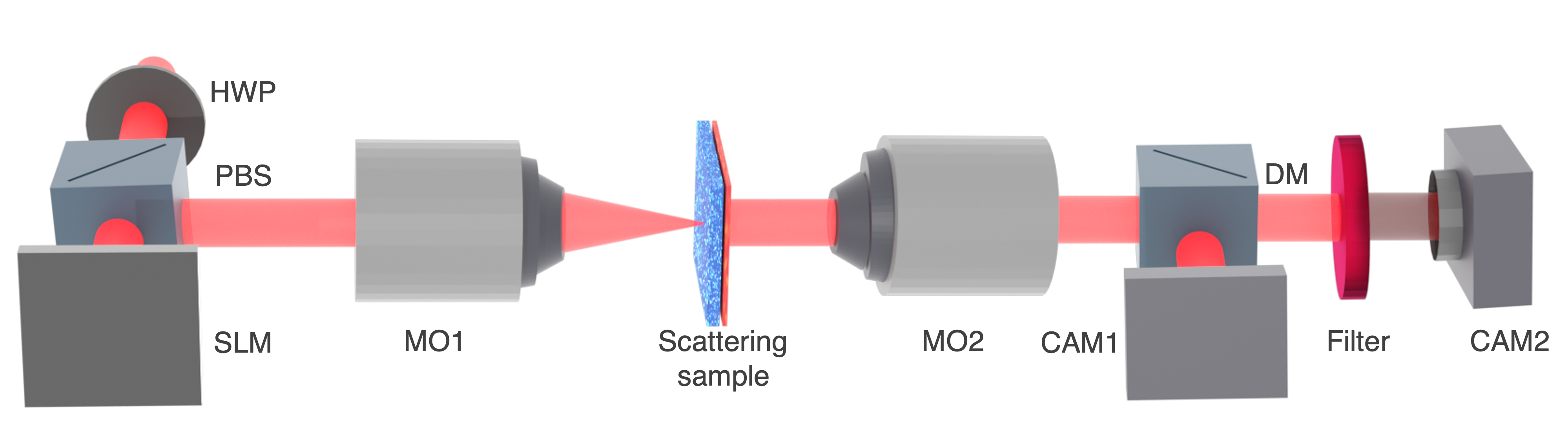}
   \caption{The experimental setup. Incident beam ($\lambda=632.8$ nm) passes through a half-wave plate (HWP) and a polarizing beam splitter (PBS) to the spatial light modulator (SLM). From the SLM, the beam is focused onto the scattering sample by a microscope objective (MO1). Speckle and fluorescence emission are collected by MO2, split by a dichroic mirror (DM), and monitored by CAM1 and CAM2, respectively. A longpass filter is added before CAM2.} 
    \label{fig:setup}
\end{figure}

The experimental optical setup is illustrated in Figure \ref{fig:setup}. The light source is a continuous wave Helium–Neon laser (HeNe, $\lambda = 632.8$ nm). The beam is expanded and incident on a phase-only SLM (Santec 200). The polarization of the incident beam is controlled by a half-wave plate (HWP) and a polarizing beam splitter (PBS). The reflected beam is transmitted through a 4$f$ imaging system and focused on the sample by a microscope objective (MO1). A second microscopic objective (MO2) is used to collect the speckle and fluorescence from the sample. A dichroic mirror (DM) splits the signals emanating from the sample. The speckle is detected by CAM1 (Thorlabs CS2100M) and the fluorescence by CAM2 (Andor Sona sCMOS). We use an additional longpass filter ($\lambda_{c}=650$ nm) to ensure CAM2 only detects fluorescence.

Plain yogurt and pig skin are used as the scattering media and the hidden objects are fluorescent microspheres (carboxylate-modified polystyrene beads, 0.04 $\mu$m  diameter). The beads are excited at  $\lambda=632.8$ nm and the emission peaks at  $\lambda=720$ nm. To prepare a 750 $\mu$m thick yogurt sample, we spread plain white yogurt on two microscope slides and sandwich the fluorescent beads between them. The pig skin sample is 2300 $\mu$m thick and is placed directly on a slide with the fluorescent beads randomly dispersed on the surface.
\section{Results and discussion}
\begin{figure}[t]
    \centering % <-- added
 \includegraphics[width=1\linewidth]{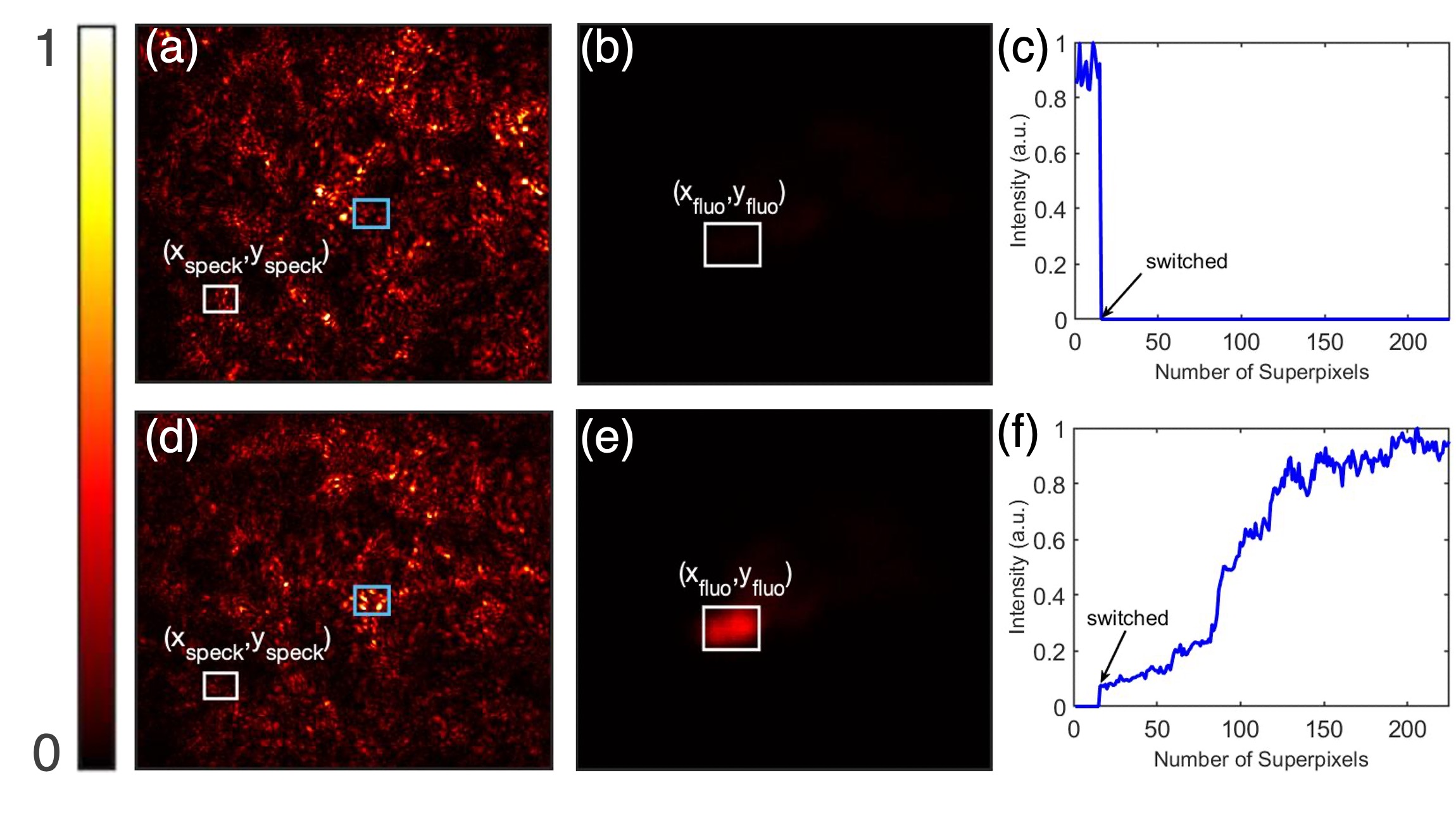}
\caption{Optimization of fluorescent signal via switching feedback (yogurt sample). (a) Initial speckle pattern, (b) fluorescence prior to optimization; (c) speckle feedback is being used initially for wavefront shaping (d) speckle and (e) fluorescence after optimization; and (f) intensity enhancement after switching to fluorescence feedback. The chosen points ($x_{speck},y_{speck}$) or ($x_{fluo},y_{fluo}$) are at the center of the white boxes and the resulting focus in the speckle is indicated by the blue rectangle.}
\label{fig:result_yogurt}
\end{figure}

Starting with case $i$), we choose a point $(x_{speck},y_{speck})$ in the speckle (CAM1) and define a target point $(x_{fluo},y_{fluo})$ in the fluorescence image (CAM2). The algorithm begins with optimizing the intensity at ($x_{speck},y_{speck}$) using the feedback from CAM1 and keeps track of the fluorescence intensity at ($x_{fluo},y_{fluo}$). The speckle pattern created by the yogurt sample (Figure \ref{fig:result_yogurt}a) has dark patches due to the absorption of the beads. A point near the darker region in the speckle is chosen as $(x_{speck},y_{speck})$, which is marked by the white rectangle.\\
Prior to the optimization process, the fluorescence emitted by the beads is negligible, as seen in Figure \ref{fig:result_yogurt}b. Once the switching condition ($I_{fluo}(target) \geq \eta \times I_{int}(target)$) is met during speckle optimization, the algorithm now switches to acquire its feedback signal from CAM2.\\
For this experiment, we set the switching parameter $\eta =1.16$. The switching condition was met after optimizing only 16 superpixels (Figure \ref{fig:result_yogurt}c). As the feedback switched to CAM2, the intensity variations at ($x_{speck},y_{speck}$) are no longer important as they do not affect the the remainder of the optimization process. But the speckle is still monitored by CAM1 and the optimization process generates a focus spot in the speckle which is indicated by the blue rectangle in Figure \ref{fig:result_yogurt}d. As seen in Figure \ref{fig:result_yogurt}e and \ref{fig:result_yogurt}f, fluorescence intensity at ($x_{fluo},y_{fluo}$) increased significantly after optimization. 
\begin{figure}[b]
    \centering % <-- added
  \includegraphics[width=1\linewidth]{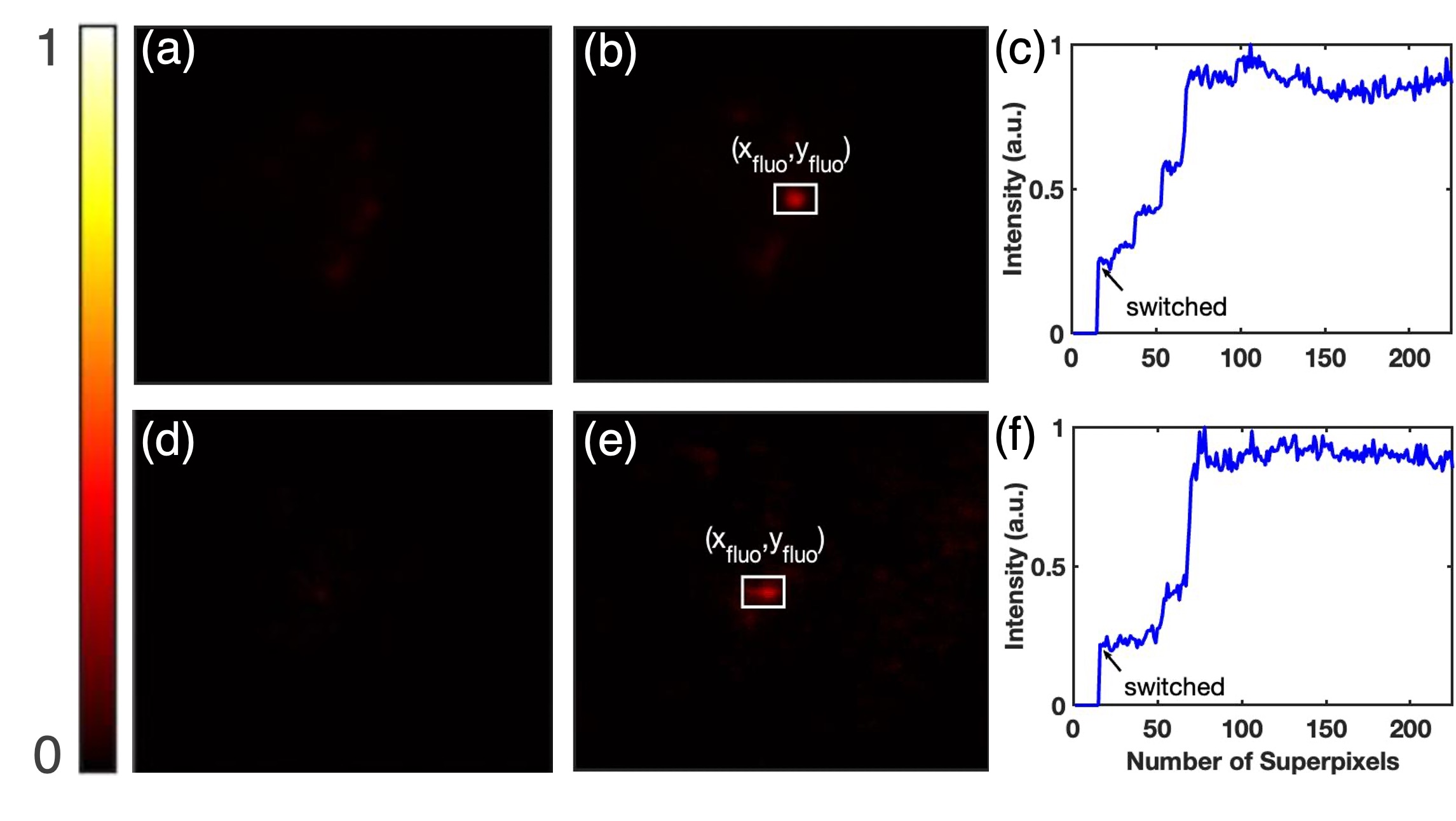}
  \label{fig:switch_skin_max}
\caption{ Fluorescence optimization at target chosen by the algorithm.
The top and bottom row have results with yogurt and pig skin sample, respectively. Fluorescence image (a) before and (b) after optimization, (c) fluorescence intensity enhancement after switching (yogurt sample); Fluorescence emitted by beads hidden by pig skin (d) before and (e) after optimization, (f) fluorescence intensity enhancement after choosing the target position and switching to fluorescence feedback; The targets picked by the algorithm are at the center of the white boxes.}
\label{fig:result_max}
\end{figure}

For case $ii$), we only define the speckle point, $(x_{speck},y_{speck})$. After optimizing the first row of superpixels with speckle feedback, the algorithm compares the initial and current fluorescence intensity for all the beads and chooses the one that exhibited the largest increase in the emitted fluorescence thus far as the target $(x_{fluo},y_{fluo})$.
Once the target is chosen the speckle optimization continues until the feedback switching condition is met.\\ 
For the data shown in Figure \ref{fig:result_max}, the top row (a$-$c) was obtained with plain yogurt as the scattering medium, and the bottom row (d$-$f) with pig skin. For both samples $\eta$ was set to 1.5. In Figure \ref{fig:result_max}a and \ref{fig:result_max}d, it can be seen that before the optimization process the fluorescence intensity is very low. After optimization, one can observe in Figure \ref{fig:result_max}b and \ref{fig:result_max}e that the intensity at the targets chosen by the algorithm have increased significantly. Despite the pig skin being thicker by an order of magnitude, the algorithm switched feedbacks after optimizing approximately the same number of superpixels (Figure \ref{fig:result_max}c and \ref{fig:result_max}f).\\
One of the benefits of case $ii$) over case $i$) is that the fluorescence optimization is to be successful even with higher values of $\eta$. Unlike case $i$) no prior knowledge of expected fluorescence position is needed for case $ii$), however, case $i$) allows one to control the target position.

Contrary to other wavefront shaping methods based on only fluorescence feedback that require higher SNR to begin the optimization process, this method is applicable even when there is no fluorescence detected initially. Additionally, this technique doesn't require any memory effect. While in most instances, the enhancement is proportional to the number of iterations\cite{yang2021}, it's not true for fluorescence optimization as increasing number of iterations will require more time and eventually result in photobleaching. In our experimental demonstration the fluorescent emission was collected in transmission geometry, however, we expect the method to work with collecting fluorescence in reflection. 

\section{Conclusion}
In summary, we have proposed a new wavefront shaping method that allows for the tracking and enhancement of initially weak signal emitted by sources hidden behind highly scattering media. We have introduced in the algorithm a smart switching function that toggles between feedback signals. We have demonstrated two main scenarios to show the flexibility of our approach. In one case the hidden target position is known, and in the other the algorithm locates the position of the hidden object and identifies the target. Our optical method is simple, easy to implement, and flexible enough to be used in either transmission or in reflection. We believe, this technique will assist in retrieving information from hidden objects inside or behind turbid media and is a stride towards the development of non-invasive optical tool for medical diagnosis. 
\vspace{-.3cm}
\section*{AUTHOR DECLARATIONS}
\vspace{-.4cm}
\subsection*{Conflict of Interest}
\vspace{-.45cm}
The authors declare no conflicts of interest.
\vspace{-.8cm}
\subsection*{Data availability}
\vspace{-.45cm}
The data that support the findings of this study are available from the corresponding author upon reasonable request.
\vspace{-.5cm}
\section*{References}
\vspace{-.5 cm}
\bibliography{switching_feedback}% Produces the bibliography via BibTeX.

\end{document}